\documentstyle[aps,prd,epsfig,preprint]{revtex}
\def\fun#1#2{\lower3.6pt\vbox{\baselineskip0pt\lineskip.9pt
\ialign{$\mathsurround=0pt#1\hfill##\hfil$\crcr#2\crcr\sim\crcr}}}
\begin{document}
\vspace{0.5in}
\title{\vskip-2.5truecm{\hfill \baselineskip 14pt 
{\hfill {{\small \hfill UT-STPD-5/98 }}}\\
{{\small \hfill THES-TP/98-05}}\vskip .1truecm} 
\vspace{1.0cm}
\vskip 0.1truecm {\bf Atmospheric Neutrino Anomaly and 
Supersymmetric Inflation }}
\vspace{1cm}
\author{{G. Lazarides}$^{(1)}$\thanks{lazaride@eng.auth.gr} 
{and N. D. Vlachos}$^{(2)}$\thanks{vlachos@physics.auth.gr}} 
\vspace{1.0cm}
\address{$^{(1)}${\it Physics Division, School of Technology, 
Aristotle University of Thessaloniki,\\ Thessaloniki GR 540 06, 
Greece.}}
\address{$^{(2)}${\it   Department of Physics, Aristotle 
University of Thessaloniki,\\Thessaloniki GR 540 06, Greece.}}
\maketitle

\vspace{2cm}

\begin{abstract}
\baselineskip 12pt

\par
A detailed investigation of hybrid inflation and the 
subsequent reheating process is performed within a 
$\mu$-problem solving supersymmetric model based on 
a left-right symmetric gauge group. The process 
of baryogenesis via leptogenesis is especially studied. For 
$\nu_{\mu}$, $\nu_{\tau}$ masses from the small angle 
MSW resolution of the solar neutrino problem and the recent 
results of the SuperKamiokande experiment, we show that 
maximal $\nu_{\mu}$-$\nu_{\tau}$ mixing can be achieved. 
The required value of the relevant coupling constant is, 
however, quite small ($\sim 10^{-6}$).

\end{abstract}

\thispagestyle{empty}
\newpage
\pagestyle{plain}
\setcounter{page}{1}
\def\beq{\begin{equation}}
\def\eeq{\end{equation}}
\def\beqa{\begin{eqnarray}}
\def\eeqa{\end{eqnarray}}
\def\tr{{\rm tr}}
\def\x{{\bf x}}
\def\p{{\bf p}}
\def\k{{\bf k}}
\def\z{{\bf z}}
\baselineskip 20pt

\par
The hybrid inflationary scenario \cite{hybrid} can be
easily implemented~\cite{lyth,dss,lss} in the context 
of supersymmetric theories in a `natural' way meaning 
that a) there is no need for tiny coupling constants, 
b) the superpotential used is the most general one 
allowed by gauge and R- symmetries, c) supersymmetry 
guarantees that radiative corrections do not invalidate 
inflation, but rather provide a slope along the 
inflationary trajectory which drives the inflaton 
towards the supersymmetric vacua, and d) supergravity 
corrections can be brought under control so as to leave 
inflation intact.

\par
A moderate extension of the minimal supersymmetric 
standard model (MSSM) based on a left-right symmetric 
gauge group provides \cite{lss} a suitable framework 
for hybrid inflation. The inflaton is associated 
with the breaking of $SU(2)_{R}$ and consists of a 
gauge singlet and a pair of $SU(2)_{R}$ doublets. 
The doublets can decay into right handed neutrinos, 
after inflation, reheating the universe and providing 
a mechanism \cite{lepto} for baryogenesis through a 
primordial leptogenesis. The gauge singlet, however, 
has no direct coupling to light matter in the simplest 
case. Moreover, its coupling to the $SU(2)_{R}$ 
doublets turns out to be unable to ensure its 
efficient decay. This difficulty can be 
overcome by introducing \cite{lss,dls} a direct 
superpotential coupling of the gauge singlet superfield 
to the electroweak higgs doublets. This way the gauge 
singlet scalar can decay into a pair of higgsinos.
It has been shown \cite{dls} that, in the presence of 
gravity-mediated supersymmetry breaking, this gauge 
singlet acquires a vacuum expectation value (vev) and 
consequently generates, through its coupling to the 
ordinary higgs superfields, the $\mu$ term of MSSM.     
A coupling of the scalar components of the $SU(2)_{R}$ 
doublets to the electroweak higgses is automatically 
induced in this scheme, allowing them to decay 
into a pair of ordinary higgses in addition to their 
useful (for baryogenesis) decay to right handed 
neutrinos.

\par
In this paper, we attempt a detailed study of 
inflation in the above scheme. In particular, we 
solve the evolution equations of this system and 
estimate the reheating temperature. The process 
of baryogenesis via leptogenesis is also 
considered and its consequences on 
$\nu_{\mu}$-$\nu_{\tau}$ mixing are analyzed. 
For masses of $\nu_{\mu}$, $\nu_{\tau}$ which 
are consistent with the small angle MSW resolution 
of the solar neutrino problem and the recent results 
of the SuperKamiokande experiment \cite{superk}, we 
examine whether maximal $\nu_{\mu}$-$\nu_{\tau}$ 
mixing can be achieved. 

\par
Let us first describe the main features of the 
$G_{LR}=SU(3)_c\times SU(2)_R\times SU(2)_L
\times U(1)_{B-L}$ symmetric model \cite{dls} 
which solves the $\mu$ problem. The $SU(2)_R
\times U(1)_{B-L}$ group is broken by a pair 
of $SU(2)_R$ doublet chiral superfields $l^c$, 
$\bar l^{c}$ which acquire a vev $M >> m_{3/2} 
\sim (0.1-1)\ {\rm TeV}$, the gravitino mass. 
This breaking is achieved by means of a gauge 
singlet chiral superfield $S$ which plays a crucial 
three-fold role: 1) it triggers $SU(2)_R$ breaking; 
2) it generates  the $\mu$ term of MSSM after 
gravity-mediated supersymmetry breaking; and 3) it 
leads to hybrid inflation \cite{hybrid}. Ignoring 
the matter fields of the model, the superpotential 
reads
\begin{equation}
W = S( \kappa l^c\bar l^{c} + \lambda h^2 - 
\kappa M^2)~, 
\label{W}
\end{equation}
where the chiral superfield $h=(h^{(1)}, h^{(2)})$ 
belongs to a bidoublet $(2,2)_{0}$ representation of
$SU(2)_L\times SU(2)_R\times U(1)_{B-L}$, and 
$h^2$ denotes the unique bilinear invariant
$\epsilon^{ij}h_i^{(1)}h_j^{(2)}$. Note that 
the parameters $\kappa$, $\lambda$ and $M$ can be 
made positive through a suitable redefinition of the 
superfields. $W$ in Eq.(\ref{W}) has the most general 
renormalizable form invariant under the gauge group 
and a continuous $U(1)$ R-symmetry under which $S$ 
carries the same charge as $W$, while $h$, $l^c$, 
$\bar l^{c}$ are neutral. This symmetry is extended 
\cite{dls} to include the matter fields of the model too 
and implies automatic baryon number conservation. It has 
been shown \cite{dls} that, after supersymmetry breaking, 
$S$ develops a vev $\langle S\rangle\approx -m_{3/2}/
\kappa$ which generates a $\mu $ term with $\mu=\lambda 
\langle S\rangle\approx -(\lambda/\kappa)m_{3/2}$.

\par
The model has a built-in inflationary trajectory in the 
field space along which the $F_S$ term is constant 
\cite{dss,lss}. This trajectory is parametrized by 
$|S|$, $|S| > S_c=M$ for $\lambda>\kappa$ (see below). 
All other fields vanish on this trajectory.   
The $F_S$ term provides us with a constant tree level 
vacuum energy density $V_{{\rm tree}}=\kappa^2 M^4$,
which is responsible for inflation. Radiative corrections 
generate a logarithmic slope \cite{dss} along the 
inflationary trajectory that drives the inflaton toward 
the minimum. The one-loop contribution to this slope comes 
from the $l^c$, $\bar l^{c}$ and $h$ supermultiplets, 
which receive at tree level a non-supersymmetric 
contribution to the masses of their scalar components from 
the $F_S$ term. For $|S| \leq S_c=M$, the $l^c$, 
$\bar l^{c}$ components become tachyonic, compensate the 
$F_S$ term and the system evolves towards the `correct' 
supersymmetric minimum at $h=0$, $l^c=\bar l^c=M$. (For 
$\kappa > \lambda$, $h$ would have become tachyonic 
earlier and the system would have evolved towards the 
`wrong' minimum at $h \neq 0$, $l^c=\bar l^c = 0$.) 
Inflation can continue at least till $|S|$ approaches 
the instability at $|S|=S_c$ provided that the slow 
roll conditions \cite{dss,lazarides} are violated only 
`infinitesimally' close to it. This is true for all values 
of the relevant parameters considered in this work. The 
cosmic microwave quadrupole anisotropy can be calculated 
\cite {dss} by standard methods and turns out to be   
\begin{equation}
\left(\frac{\delta T}{T}\right)_{Q} \approx 
\frac{32 \pi^{5/2}}
{3\sqrt{5}}\left(\frac{M}{M_P}\right)^{3}
\kappa^{-1}x_{Q}^{-1}\Lambda (x_{Q})^{-1}~,
\label{anisotropy}
\end{equation}
where $M_{P}=1.22\times 10^{19}~{\rm{GeV}}$ is the 
Planck scale and
\begin{eqnarray*}
\Lambda(x)=\left(\frac{\lambda}
{\kappa}\right)^{3}\left[\left(\frac{\lambda}
{\kappa}x^2-1\right)\ln \left(1-\frac{\kappa}
{\lambda}x^{-2}\right)
+\left(\frac{\lambda}{\kappa}x^2+1\right)
\ln \left(1+\frac{\kappa}{\lambda}x^{-2}\right)\right] 
\end{eqnarray*}
\begin{equation}
+(x^2-1)\ln (1-x^{-2})+(x^2+1)\ln (1+x^{-2})~,
\label{temp}
\end{equation}
with $x=|S|/S_c$ and $S_Q$ being the value of $|S|$ when 
the present horizon scale crossed outside the inflationary 
horizon. The number of e-foldings experienced by the 
universe between the time the quadrupole scale exited the 
horizon and the end of inflation is 
\begin{equation} 
N_Q \approx 32 \pi^3 \left(\frac{M}{M_P}\right)^2 
\kappa^{-2}\int_{1}^{x_{Q}^{2}}\frac{dx^2}{x^2}
\Lambda(x)^{-1}~.
\label{efoldings}
\end{equation}
The spectral index of density perturbations turns out to 
be very close to unity.

\par
After reaching the instability at $|S|=S_c$, the system 
undergoes \cite{bl} a short complicated evolution during 
which inflation continues for another e-folding or so. 
The energy density of the system is reduced by a factor 
of about 2-3 during this period. The system then rapidly 
settles in a regular oscillatory phase about the 
supersymmetric vacuum. Parametric resonance can be ignored 
in this case \cite{bl}. The inflaton (oscillating system) 
consists of the two complex scalar fields $S$ and 
$\theta=(\delta \phi + \delta\bar{\phi})/\sqrt{2}$, 
where $\delta \phi = \phi - M$, 
$\delta \bar{\phi} = \bar{\phi} - M$, with mass 
$m_{infl}=\sqrt{2}\kappa M$. 
Here $\phi$, $\bar{\phi}$ are the neutral components of 
the superfields $l^c$, $\bar l^{c}$ respectively. The 
scalar fields $S$ and $\theta$ predominantly decay into 
ordinary higgsinos and higgses respectively with a common 
decay width $\Gamma_{h}=(1/16\pi)\lambda^{2}m_{infl}$, 
as one can easily deduce from the couplings in Eq.(\ref{W}).
Note, however, that $\theta$ can also decay to right 
handed neutrinos $\nu^c$ through the non-renormalizable 
superpotential term $(M_{\nu^c}/2M^{2})\bar{\phi} 
\bar{\phi} \nu^c \nu^c$, allowed by the gauge and R- 
symmetries of the model \cite{lss}. Here, $M_{\nu^c}$ 
denotes the Majorana mass of the relevant $\nu^c$. The 
scalar $\theta$ decays preferably into the heaviest 
$\nu^c$ with $M_{\nu^{c}} \leq m_{infl}/2$. The 
decay rate is given by
\begin{equation}
\Gamma_{\nu^c} \approx \frac{1}{16\pi}~\kappa^2 
m_{infl}~ \alpha^2 (1-\alpha^2)^{1/2}~,
\label{decayneu}
\end{equation}
where $0\leq \alpha=2M_{\nu^c}/m_{infl} \leq 1$.
The subsequent decay of these $\nu^{c}$ 's gives 
rise to a primordial lepton number \cite{lepto}. The 
baryon asymmetry of the universe can then be obtained 
by partial conversion of this lepton asymmetry through 
sphaleron effects.
 
\par
The energy densities $\rho_{S}$, $\rho_{\theta}$, 
and $\rho_{r}$ of the oscillating fields $S$, $\theta$, 
and the `new' radiation produced by their decay to 
higgsinos, higgses and $\nu^c$ 's are controlled by the 
equations:
\begin{equation}
\dot{\rho}_{S}=-(3H+\Gamma_{h})\rho_{S}~,
~\dot{\rho}_{\theta}=-(3H+\Gamma_{h}+
\Gamma_{\nu^{c}})\rho_{\theta}~,
\label{infdensity}
\end{equation}
\begin{equation}
\dot{\rho}_{r}=-4H\rho_{r}+\Gamma_{h}\rho_{S}+
(\Gamma_{h}+\Gamma_{\nu^{c}})\rho_{\theta}~,
\label{raddensity}
\end{equation}
where
\begin{equation}
H=\frac{\sqrt{8\pi}}{\sqrt{3}M_P}~(\rho_{S}+
\rho_{\theta}+\rho_{r})^{1/2}~,
\label{hubble}
\end{equation}
is the Hubble parameter and overdots denote derivatives 
with respect to cosmic time $t$. We have assumed that the 
potential energy density is, to a good approximation, 
quadratic in the fields $S$ and $\theta$ and, thus, the 
oscillating inflaton system resembles the behavior of 
`matter'. Note that the second equation in 
Eq.(\ref{infdensity}) can be replaced by
\begin{equation}
\rho_{\theta}(t)=\rho_{S}(t)
e^{-\Gamma_{\nu^c}(t-t_0)}~,
\label{rhotheta}
\end{equation}
where $t_0$ is the cosmic time at the onset of the 
oscillatory phase. The initial values are taken to be 
$\rho_{S}(t_0)=\rho_{\theta}(t_0)
\approx \kappa^{2}M^{4}/6$, $\rho_{r}(t_{0})=0$ and, 
for all practical purposes, we put $t_0=0$. The `reheat' 
temperature $T_{r}$ is calculated from the equation
\begin{equation}
\rho_{S}+\rho_{\theta}=\rho_{r}=
\frac{\pi^2}{30}~g_{*}T_{r}^{4}~,
\label{reheat}
\end{equation}
where the effective number of massless degrees of 
freedom is $g_{*}$=228.75 for MSSM.

\par
The lepton number density $n_{L}$ produced by the 
$\nu^{c}$ 's satisfies the evolution equation:
\begin{equation}
\dot{n}_{L}=-3Hn_{L}+2\epsilon\Gamma_{\nu^c}
n_{\theta}~,
\label{lepton}
\end{equation}
where $\epsilon$ is the lepton number produced per 
decaying right handed neutrino and the factor of 2 in 
the second term of the rhs comes from the fact that we
get two $\nu^c$ 's for each decaying scalar $\theta$ 
particle. Eq.(\ref{lepton}) is easily integrated out 
to give
\begin{equation}
n_{L}(t)\sim n_{\theta}(t_0)
\left(\frac{a(t)}{a(t_0)}\right)^{-3}
\frac{2\epsilon \Gamma_{\nu^c}}
{\Gamma_{h}+\Gamma_{\nu^c}}~, 
~~\rm{as}~~t\rightarrow \infty~,
\label{leptonasymptotic}
\end{equation}
where $a(t)$ is the scale factor of the universe. The 
first equation in Eq.(\ref{infdensity}) gives
\begin{equation}
\rho_{S}(t)=\rho_{S}(t_0)\left(\frac{a(t)}{a(t_0)}
\right)^{-3}e^{-\Gamma_{h}(t-t_0)}~.
\label{rhoes}
\end{equation}
Combining Eqs.(\ref{leptonasymptotic}) and (\ref{rhoes}) 
we get the asymptotic ($t\rightarrow\infty$) lepton 
asymmetry
\begin{equation}
\frac{n_{L}(t)}{s(t)}\sim 3
\left(\frac{15}{8}\right)^{1/4}\pi^{-1/2}
g_{*}^{-1/4}m_{infl}^{-1}~
\frac{\epsilon\Gamma_{\nu^c}}
{\Gamma_{h}+\Gamma_{\nu^c}}~\rho_{r}^{-3/4}
\rho_{S}e^{\Gamma_{h}t}~,
\label{leptonasymmetry}
\end{equation}
where
\begin{equation}
s(t)\sim \frac{2\pi^{2}g_{*}}{45}
\left(\frac{30}{\pi^2g_{*}}\right)^{3/4}
\rho_{r}^{3/4}
\label{entropy}
\end{equation}
is the asymptotic entropy density. For MSSM spectrum
between $100$ GeV and $M$, the observed baryon 
asymmetry $n_{B}/s$ is related \cite{ibanez} to 
$n_{L}/s$ by $n_{B}/s=-(28/79)(n_{L}/s)$. 
It is important to ensure that the primordial lepton 
asymmetry is not erased by lepton number violating 
$2\rightarrow 2$ scatterings at all temperatures 
between $T_r$ and 100 GeV. This requirement gives 
\cite{ibanez} $m_{\nu_{\tau}}\stackrel{_<}{_\sim} 
10~{\rm{eV}}$ which is readily satisfied in our case 
(see below).

\par
Assuming hierarchical light neutrino masses, we take 
$m_{\nu_{\mu}}\approx 2.6\times 
10^{-3}~\rm{eV}$ which is the central value of the 
$\mu$-neutrino mass coming from the small 
angle MSW resolution of the solar neutrino 
problem \cite{smirnov}. The $\tau$-neutrino mass
will be restricted by the atmospheric anomaly 
\cite{superk} in the range 
$3\times 10^{-2}~\rm{eV}\stackrel{_{<}}{_{\sim }}
m_{\nu _{\tau }}\stackrel{_{<}}{_{\sim }}11\times 
10^{-2}~\rm{eV}$. Recent analysis \cite{giunti} of the 
results of the CHOOZ experiment \cite{chooz} shows that 
the oscillations of solar and atmospheric neutrinos 
decouple. We thus concentrate on the two heaviest families 
ignoring the first one. Under these circumstances, the 
lepton number generated per decaying $\nu^c$ 
is \cite{lazarides,neu}
\begin{equation}
\epsilon=\frac{1}{8\pi}~g
\left(\frac{M_3}{M_2}\right)
~\frac{{\rm c}^{2}{\rm s}^{2}\ 
\sin 2\delta \ 
(m_{3}^{D}\,^{2}-m_{2}^{D}\,^{2})^{2}}
{|\langle h^{(1)}\rangle|^{2}~(m_{3}^{D}\,^{2}\ 
{\rm s}^{2}\ +
\ m_{2}^{D}\,^{2}{\rm \ c^{2}})}~,
\label{epsilon}
\end{equation}
where $g(r)=r\ln (1 + r^{-2})$~, 
$|\langle h^{(1)}\rangle|\approx 174~\rm{GeV}$, 
${\rm c}=\cos \theta ,\ {\rm s}=\sin \theta $, 
and $\theta$ ($0\leq \theta \leq \pi /2$) and 
$\delta$ ($-\pi/2\leq \delta <\pi/2 $) are the 
rotation angle and phase which diagonalize the Majorana 
mass matrix of $\nu^{c}$ 's, $M^{R}$, with eigenvalues 
$M_2$, $M_3$ ($\geq 0$) in the basis where the `Dirac' 
mass matrix of the neutrinos, $M^{D}$, is diagonal with 
eigenvalues $m_{2}^{D}$, $m_{3}^{D}$ ($\geq 0$). Note 
that, for the range of parameters considered here, the 
scalar $\theta$ decays into the second heaviest right 
handed neutrino with mass $M_{2}$ ($<M_{3}$) and, thus, 
$M_{\nu^{c}}$ in Eq.(\ref{decayneu}) should be 
identified with $M_{2}$. Moreover, $M_{3}$ turns out 
to be bigger than $m_{infl}/2$ as it should. We will 
denote the two positive eigenvalues of the light neutrino 
mass matrix by $m_{2}$ (=$m_{\nu _{\mu }}$), $m_{3}$ 
(=$m_{\nu _{\tau }}$) with $m_{2}\leq m_{3}$. All the 
quantities here (masses, rotation angles and phases) are 
`asymptotic' (defined at the grand unification scale 
$M_{GUT}$). The determinant and the trace invariance of 
the light neutrino mass matrix imply\cite{neu} 
two constraints on the (asymptotic) parameters which 
take the form: 
\begin{equation}
m_{2}m_{3}\ =\ \frac{\left( m_{2}^{D}m_{3}^{D}
\right) ^{2}}{M_{2}\ M_{3}}~,
\label{determinant}
\end{equation}
\begin{eqnarray*}
m_{2}\,^{2}+m_{3}\,^{2}\ =\frac{\left( m_{2}^{D}\,\,
^{2}{\rm c}^{2}+m_{3}^{D}\,^{2}{\rm s}^{2}\right) 
^{2}}{M_{2}\,^{2}}+
\end{eqnarray*}
\begin{equation}
\ \frac{\left( m_{3}^{D}\,^{2}{\rm c}^{2}+m_{2}^{D}\,
^{2}{\rm s}^{2}\right)^{2}}{M_{3}\,^{2}}+\ 
\frac{2(m_{3}^{D}\,^{2}-m_{2}^{D}\,^{2})^{2}
{\rm c}^{2}{\rm s}^{2}\,{\cos 2\delta }}
{M_{2}\,M_{3}}~\cdot
\label{trace} 
\end{equation}

\par
The $\mu$-$\tau$ mixing angle $\theta _{23}$ 
(=$\theta _{\mu\tau}$) lies \cite{neu} in the range
\begin{eqnarray*}
|\,\varphi -\theta ^{D}|\leq \theta _{23}\leq
\varphi +\theta ^{D},\ {\rm {for}\ \varphi +
\theta }^{D}\leq \ \pi /2~,~~~~~
\end{eqnarray*}
\begin{equation}
|\,\varphi -\theta ^{D}|\leq \theta _{23}\leq
\pi-\varphi -\theta ^{D},\ {\rm {for}\ \varphi +
\theta }^{D}\geq \ \pi /2~,
\label{mixing}
\end{equation}
where $\varphi$ ($0\leq \varphi \leq \pi /2$) is the 
rotation angle which diagonalizes the light neutrino mass 
matrix, $m=-\tilde{M}^{D}{M^{R}}^{-1} M^{D}$, in the 
basis where the `Dirac' mass matrix is diagonal 
and $\theta ^{D}$ ($0\leq \theta ^{D} \leq \pi /2$) 
is the `Dirac' mixing angle in the 2-3 leptonic sector 
(i.e., the `unphysical' leptonic mixing angle in the 
absence of the Majorana masses of the $\nu^{c}$ 's).

\par
The `asymptotic' Dirac masses of $\nu_{\mu}$, 
$\nu_{\tau}$ as well as $\theta ^{D}$ can be 
related to the quark sector parameters by assuming 
approximate $SU(4)_{c}$ symmetry. We obtain the 
asymptotic relations: $m_{2}^{D}\approx m_{c}\ ,
\ m_{3}^{D}\approx \ m_{t}\ ,
\ \sin\theta ^{D}\approx |V_{cb}|$~.
Renormalization effects must now be taken into account. 
To this end, we take MSSM spectrum and large 
$\tan \beta \approx m_{t}/m_{b}$. The latter follows 
from the fact that the MSSM higgs doublets form a 
$SU(2)_{R}$ doublet. It turns out \cite{neu} that, 
in this case, renormalization effects can be accounted 
for by simply substituting in the above formulae the 
following numerical values: $m_{2}^{D}\approx 0.23~
{\rm GeV}$, $\ m_{3}^{D}\approx 116$ GeV and 
$\sin \theta ^{D}\approx 0.03$. Also, $\tan^{2} 2 
\theta _{23}$ increases by about 40\% from 
$M_{GUT}$ to $M_{Z}$.

\par
\begin{figure}
\epsfig{figure=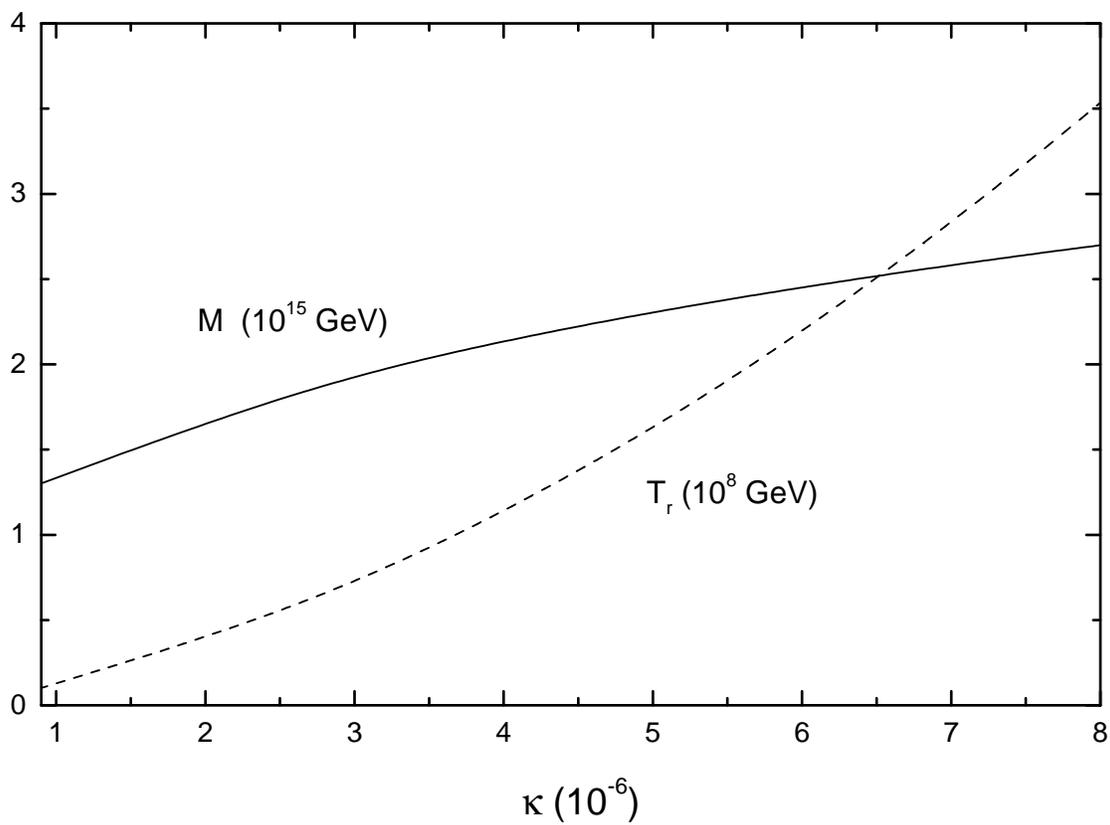,height=5.8in,angle=-90}
\medskip
\caption{The mass scale $M$ (solid line) and the reheat 
temperature $T_{r}$ (dashed line) as functions of 
$\kappa$.
\label{reheating}}
\end{figure}

In order to predict the $\nu_{\mu}$-$\nu_{\tau}$ 
mixing, we take a specific MSSM framework \cite{als} 
where the three Yukawa couplings of the third generation 
unify `asymptotically' and, consequently, 
$\tan \beta \approx m_{t}/m_{b}$. We choose the 
universal scalar mass (gravitino mass) $m_{3/2} 
\approx 290~{\rm{GeV}}$ and the universal gaugino mass 
$M_{1/2} \approx 470~{\rm{GeV}}$. These values correspond 
\cite{asw} to $m_{t}(m_{t})\approx 166~{\rm{GeV}}$ and 
$m_{A}$ (the tree level mass of the CP-odd scalar higgs 
boson) =$M_{Z}$. The `asymptotic' higgsino mass $\mu$ is 
related \cite{carena} to these parameters by
$|\mu|/m_{3/2}\approx (M_{1/2}/m_{3/2})
(1-Y_{t}/Y_{f})^{-3/7}$, where $Y_{t}=h_{t}^2
\approx 0.91$ is the square of the top-quark Yukawa 
coupling and $Y_{f}\approx 1.04$ is the weak scale 
value of $Y_{t}$ corresponding to `infinite' value at 
$M_{GUT}$. For these numerical values, we obtain
$\lambda/\kappa=|\mu|/m_{3/2}\approx 3.95$ which can be 
substituted in Eqs.(\ref{anisotropy})-(\ref{efoldings}).
These equations can then be solved, for
$(\delta T/T)_{Q} \approx 6.6\times 10^{-6}$ from COBE, 
$N_Q \approx 50$ and any value of $x_{Q}>1$. 
Eliminating $x_{Q}$, we obtain $M$ as a function of 
$\kappa$ depicted in Fig.\ref{reheating}. The evolution 
equations (\ref{infdensity})-(\ref{hubble}) are solved 
numerically for each value of $\kappa$. The parameter 
$\alpha^{2}$ in Eq.(\ref{decayneu}) is taken to be equal 
to 2/3. This choice maximizes the decay width, 
$\Gamma_{\nu^{c}}$, of the inflaton to $\nu^{c}$ 's and, 
thus, the subsequently produced primordial lepton asymmetry. 
The reheat temperature, $T_{r}$, is then calculated from 
Eq.(\ref{reheat}) for each value of $\kappa$. The result is 
again depicted in Fig.\ref{reheating}.

\par
\begin{figure}
\epsfig{figure=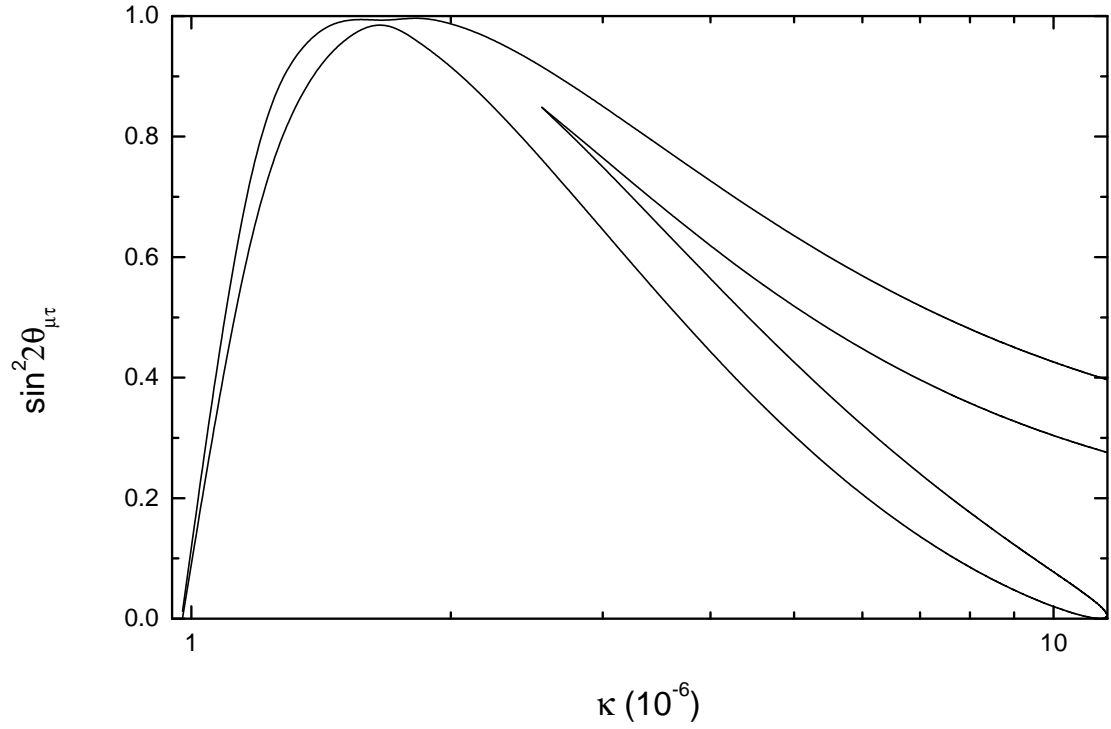,height=5.8in,angle=-90}
\medskip
\caption{The allowed region (bounded by the solid lines)
in the $\kappa$-$\sin^{2} 2 \theta _{\mu\tau}$ plane 
for $m_{\nu_{\mu}}\approx 2.6\times 10^{-3}~\rm{eV}$ 
and $m_{\nu_{\tau}}\approx 7 \times 10^{-2}~\rm{eV}$.
\label{angle}}
\end{figure}

We next evaluate the lepton asymmetry. We begin by first 
considering the central value of $m_{\nu_{\tau}}\approx 7 
\times 10^{-2}~\rm{eV}$ given by the SuperKamiokande 
experiment \cite{superk} (the $\mu$-neutrino mass is kept 
fixed to its central value $m_{\nu_{\mu}}\approx 2.6
\times 10^{-3}~\rm{eV}$ from the MSW resolution of the 
solar neutrino puzzle). The mass of the second heaviest 
$\nu^c$, into which the scalar $\theta$ decays partially, 
is given by $M_{2}=M_{\nu^{c}}=\alpha m_{infl}/2$ and 
$M_{3}$ can be found from Eq.(\ref{determinant}). 
We can use the trace condition in Eq.(\ref{trace}) to 
solve for $\delta(\theta )$ in the interval 
$0 \leq \theta \leq \pi /2$. The expression for 
$\delta (\theta)$ is subsequently substituted in 
Eq.(\ref{epsilon}) for $\epsilon$. The leptonic 
asymmetry as a function of the angle $\theta$ can be found 
from Eq.(\ref{leptonasymmetry}). To each value of 
$\kappa$ correspond two values of the angle $\theta$ 
satisfying the low deuterium abundance constraint 
$\Omega _{B}h^{2}\approx 0.025$. (These values of 
$\theta$ turn out to be quite insensitive to the exact 
value of $n_{B}/s$.) The corresponding values of the 
rotation angle $\varphi$, which diagonalizes the light 
neutrino mass matrix, are then found and the allowed 
region of the mixing angle $\theta _{\mu\tau}$ 
in Eq.(\ref{mixing}) is determined. Taking into 
account renomalization effects and superimposing all the 
permitted regions, we obtain the allowed range of $\sin^{2} 
2 \theta _{\mu\tau}$ as a function of $\kappa$, shown 
in Fig.\ref{angle}. We observe that maximal mixing
($\sin^{2} 2 \theta _{\mu\tau}\approx 1$) is achieved 
for $1.5\times 10^{-6}\stackrel{_{<}}{_{\sim }}\kappa
\stackrel{_{<}}{_{\sim }}1.8\times 10^{-6}$. Also, 
$\sin^{2} 2 \theta _{\mu\tau}\stackrel{_{>}}{_{\sim }} 
0.8$ \cite{superk} corresponds to $1.2\times 10^{-6}
\stackrel{_{<}}{_{\sim }}\kappa\stackrel{_{<}}
{_{\sim }}3.4\times 10^{-6}$. 

\par
\begin{figure}
\epsfig{figure=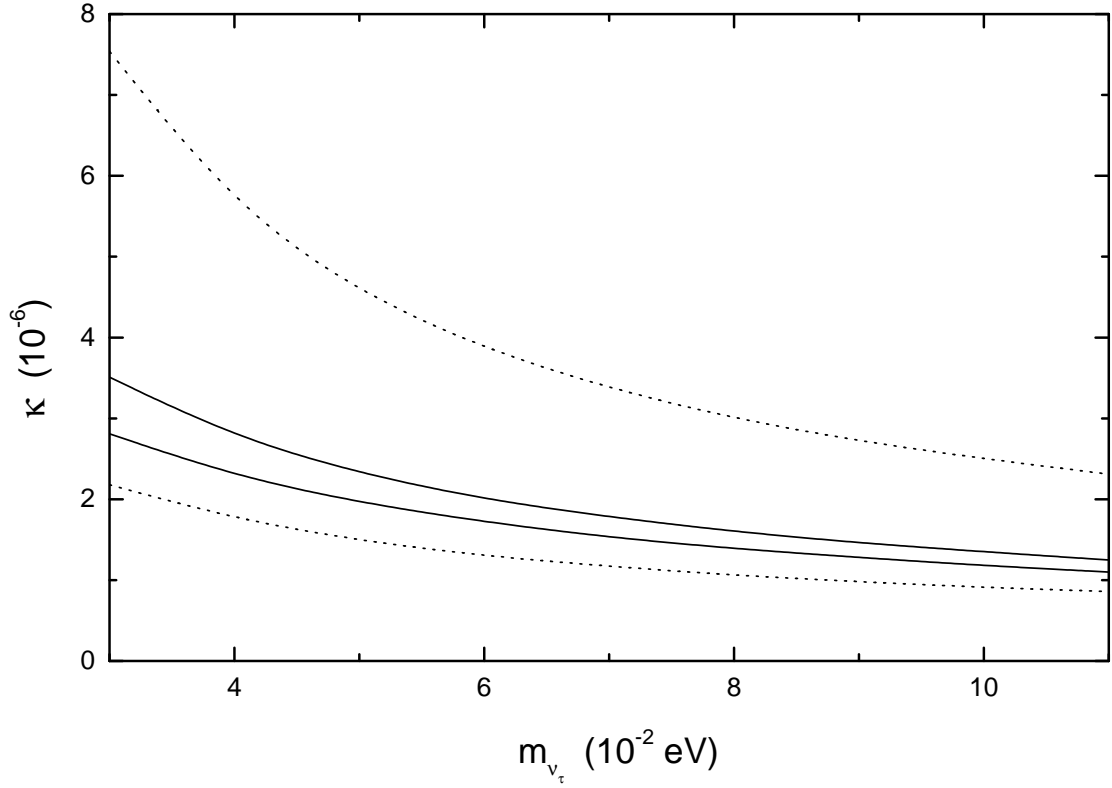,height=5.8in,angle=-90}
\medskip
\caption{The regions on the $m_{\nu_{\tau}}$-$\kappa$
plane corresponding to maximal $\nu_{\mu}$-$\nu_{\tau}$ 
mixing (bounded by the solid lines) and 
$\sin^{2} 2 \theta _{\mu\tau}
\stackrel{_{>}}{_{\sim }} 0.8$ (bounded by the dotted 
lines). Here we consider the range $3\times 
10^{-2}~\rm{eV}\stackrel{_{<}}{_{\sim }}
m_{\nu _{\tau }}\stackrel{_{<}}{_{\sim }}
11\times 10^{-2}~\rm{eV}$ ($m_{\nu_{\mu}}
\approx 2.6\times 10^{-3}~\rm{eV}$).
\label{kappa}}
\end{figure}

The analysis above can be repeated for all values of
$m_{\nu _{\tau }}$ allowed by SuperKamiokande. 
The allowed regions in the $m_{\nu _{\tau }}$-$\kappa$ 
plane for maximal $\nu_{\mu}$-$\nu_{\tau}$ mixing 
(bounded by the solid lines) and $\sin^{2} 2 
\theta _{\mu\tau}\stackrel{_{>}}{_{\sim }} 0.8$ 
(bounded by the dotted lines) are shown in Fig.\ref{kappa}. 
Notice that, for $\sin^{2} 2 
\theta _{\mu\tau}\stackrel{_{>}}{_{\sim }} 0.8$, 
$\kappa\approx(0.9-7.5)\times 10^{-6}$ which is rather 
small. (Fortunately, supersymmetry protects this
coupling from radiative corrections.) The corresponding
values of $M$ and $T_{r}$ can be read from 
Fig.\ref{reheating}. We find $1.3\times 
10^{15}~{\rm{GeV}}\stackrel{_{<}}{_{\sim }} 
M\stackrel{_{<}}{_{\sim }}
2.7\times 10^{15}~{\rm{GeV}}$ and $10^{7}
~{\rm{GeV}}\stackrel{_{<}}{_{\sim }} T_{r}
\stackrel{_{<}}{_{\sim }}3.2\times 
10^{8}~{\rm{GeV}}$. We observe that $M$ turns out 
to be somewhat smaller than the MSSM unification scale 
$M_{GUT}$. (It is anticipated that $G_{LR}$ is embedded 
in a grand unified theory.) The reheat temperature, 
however, satisfies the gravitino constraint 
($T_{r}\stackrel{_{<}}{_{\sim}} 10^{9}~{\rm{GeV}}$).
It should be noted that, for the values of the parameters
chosen here, the lightest supersymmetric particle (LSP)
is \cite{asw} an almost pure bino with mass $m_{LSP}
\approx 0.43M_{1/2}\approx 200~{\rm{GeV}}$ 
\cite{drees}. Its contribution to the mass of the universe 
turns out \cite{drees} to be 
$\Omega_{LSP}h^{2}\approx 1$.

\par
In summary, we have investigated hybrid inflation and 
the subsequent reheating process in the framework of a
$\mu$-problem solving supersymmetric model based on a 
left-right symmetric gauge group. The process of 
baryogenesis via leptogenesis is especially considered. 
For masses of $\nu_{\mu}$, $\nu_{\tau}$ consistent 
with the small angle MSW resolution of the solar neutrino 
problem and the recent SuperKamiokande data, we showed 
that maximal $\nu_{\mu}$-$\nu_{\tau}$ mixing can 
be achieved. The required value of the coupling constant 
$\kappa$ is, however, quite small ($\sim 10^{-6}$).
 
\vspace{0.5cm}
We would like to thank B. Ananthanarayan and C. Pallis 
for useful discussions. This work was supported by the 
research grant PENED/95 K.A.1795.

\def\ijmp#1#2#3{{ Int. Jour. Mod. Phys. }{\bf #1~}(19#2)~#3}
\def\pl#1#2#3{{ Phys. Lett. }{\bf B#1~}(19#2)~#3}
\def\zp#1#2#3{{ Z. Phys. }{\bf C#1~}(19#2)~#3}
\def\prl#1#2#3{{ Phys. Rev. Lett. }{\bf #1~}(19#2)~#3}
\def\rmp#1#2#3{{ Rev. Mod. Phys. }{\bf #1~}(19#2)~#3}
\def\prep#1#2#3{{ Phys. Rep. }{\bf #1~}(19#2)~#3}
\def\pr#1#2#3{{ Phys. Rev. }{\bf D#1~}(19#2)~#3}
\def\np#1#2#3{{ Nucl. Phys. }{\bf B#1~}(19#2)~#3}
\def\mpl#1#2#3{{ Mod. Phys. Lett. }{\bf #1~}(19#2)~#3}
\def\arnps#1#2#3{{ Annu. Rev. Nucl. Part. Sci. }{\bf
#1~}(19#2)~#3}
\def\sjnp#1#2#3{{ Sov. J. Nucl. Phys. }{\bf #1~}(19#2)~#3}
\def\jetp#1#2#3{{ JETP Lett. }{\bf #1~}(19#2)~#3}
\def\app#1#2#3{{ Acta Phys. Polon. }{\bf #1~}(19#2)~#3}
\def\rnc#1#2#3{{ Riv. Nuovo Cim. }{\bf #1~}(19#2)~#3}
\def\ap#1#2#3{{ Ann. Phys. }{\bf #1~}(19#2)~#3}
\def\ptp#1#2#3{{ Prog. Theor. Phys. }{\bf #1~}(19#2)~#3}
\def\plb#1#2#3{{ Phys. Lett. }{\bf#1B~}(19#2)~#3}

\newpage

\pagestyle{empty}

\end{document}